\documentstyle[12pt]{article}
\begin{document}

\title{On Bifurcation Points of the Stationary Vlasov-Maxwell System
with Bifurcation Direction}

\author{{\Large\bf N.A.Sidorov $^{(*)}$, A.V.Sinitsyn $^{(**)}$}\\ [0.9cm]
$^{(*)}$ Department of Mathematics\\
Irkutsk State University\\
Karl-Marks Str., 1\\
Irkutsk 664003, Russia,\\
e-mail: sid@math.isu.runnet.ru   \\
$^{(**)}$ Institute Dynamics of Systems\\
and Control Theory\\
P.O.Box 1233\\
Irkutsk 664033, Russia,\\
e-mail: avsin@icc.ru
}

\date{}
\thispagestyle{empty}
\maketitle

Let us consider a multicomponent plasma containing electrons and positively
charged ions of various kinds and described by the multiparticle distribution
function $f_{i}=f_{i}(r,v)$, $i=1,\ldots,N$. Plasma occupies a domain
$D\subset R^{3}$ with smooth boundary. The particles interact via a
self-consistent force fields; collisions of particles are neglected.

The plasma is described by the following version of a nonrelativistic stationary
Vlasov-Maxwell(VM) system [1]
$$
v\cdot\partial_{r}f_{i}+q_{i}/m_{i}(E+\frac {1} {c}v\times B)
\cdot\partial_{v}f_{i}=0
$$
$$
r\in D\subset R^{3}, \;\;v\in R^{3},\;\;\;i=1,\ldots,N,
$$
$$
curl E=0,
$$
$$
div B=0,    \eqno(I)
$$
$$
div E=4\pi\sum_{k=1}^{N}q_{k}\int_{R^{3}}f_{k}(r,v){\rm d}v\stackrel{\triangle}
{=}\rho,
$$
$$
curl B=\frac{4\pi}{c}\sum_{k=1}^{N}q_{k}\int_{R^{3}}vf_{k}(r,v){\rm d}v\stackrel
{\triangle}{=}j.
$$
Here $\rho(r)$, $j(r)$ are the densities of the charge and the current,
respectively, and $E(r)$, $B(r)$ are the electric and the magnetic field.

We shall search the solution $E,B,f$ of VM system (I) for $r\in D\subset R^{3}$
with the boundary conditions on the potentials and the densities
$$
U\mid_{\partial D}=u_{01}, \;\;\;\;(A,d)\mid_{\partial D}=u_{02}, \eqno(2)
$$
$$
\rho\mid_{\partial D}=0, \;\;\;\;j\mid_{\partial D}=0, \eqno(3)
$$
where $E=-\partial_{r}U$, $B=curl A$, and $U, A$ are the scalar and the
vector potentials.

The solution $E^{0}$, $B^{0}$, for which $\rho^{0}=0$, $j^{0}=0$ in a domain
$D$, will be referred to as the trivial one.

Our aim is to obtain the existence theorem of nontrivial solutions for the
boundary-value problem (I), (2), (3).

In this paper we restrict ourselves to distributions of the form
$$
f_{i}(\lambda,r,v)=a(\lambda)\hat{f_{i}}(-\alpha_{i}(\lambda)v^{2}+
\varphi_{i}(\lambda,r), \;v\cdot d_{i}(\lambda)+\psi_{i}(\lambda,r))
\stackrel{\triangle}{=}a(\lambda)\hat{f_{i}}(\lambda,{\bf R}, {\bf G}), \eqno(4)
$$
$$
\varphi_{i}:\; R^{3}\longrightarrow R:\;\;\; \psi_{i}: \; R^{3}\longrightarrow R, \;\;\; r\in D
\subseteq R^{3}, \;\;\; v\in R^{3};
$$
$$
\lambda\in R,\;\;\;a\in R^{+},\;\;\; \alpha_{i}\in R^{+}\stackrel{\triangle}{=}[0,\infty),;\;\;\;
d_{i}\in R^{3},\;\;\;i=1,\ldots,N,
$$
where the functions $\varphi_{i}$, $\psi_{i}$ generating the corresponding electromagnetic field $
(E,\;B)$ are to be determined. Let us note that similar distribution functions
were introduced in [2] and have been used in [3].
We are interested in the dependence of desired functions $\varphi_{i}$,
$\psi_{i}$ generating the nontrivial solutions on the parameter $\lambda$
in distribution (4). The case, when the parameters $\alpha_{i}$, $d_{i}$ are
independent of $\lambda$ was considered in [4-6]. Here we consider the general case.

Let us give a one preliminary result on reduction of the VM system (I) with
conditions (2) to the qusilinear system of elliptic equations for distribution (4)
[7].
We assume  that the following condition is satisfied:

{\bf A.} $\hat{f_{i}}$({\bf R}, {\bf G}) in (4) are fixed, differentiable functions;
$\alpha_{i}$, $d_{i}$ are free parameters, $\mid d_{i}\mid\neq 0$; $\varphi_{i}=c_{1i}+l_{i}\varphi(r)$,
$\psi_{i}=c_{2i}+k_{i}\psi(r)$, $c_{1i}$, $c_{2i}$ are constants;
the parameters $l_{i}$, $k_{i}$ are related by equations
$$
l_{i}=\frac {m_{1}}{\alpha_{1}q_{i}} \frac {\alpha_{i}q_{i}}{m_{i}}, \;\;\;\;\; k_{i}\frac {q_{1}}{m_{1}}d_{1}=
\frac {q_{i}}{m_{i}}d_{i};
$$
the integrals
$$
\int_{R^{3}}\hat{f_{i}}\;{\rm d}v, \;\;\;\; \int_{R^{3}}\hat{f_{i}}v\;{\rm d}v
$$
converge for all $\varphi_{i}$, $\psi_{i}$.

{\bf Theorem 1.} {\it Suppose that the distribution function $f_{i}$ has the form
(4) and  condition {\bf A} is satisfied. Let the
vector-function $(\varphi,\;\psi)$ be a solution of system of equations
$$
\bigtriangleup\varphi=a(\lambda)\mu\sum_{k=1}^{N}q_{k}\int_{R^{3}}f_{k}(\lambda)
{\rm d}v, \;\;\;\;\mu=\frac{8\pi\alpha q}{m},
$$
$$
\eqno(5)
$$
$$
\bigtriangleup\psi=a(\lambda)\nu\sum_{k=1}^{N}q_{k}\int_{R^{3}}(v,d)f_{k}(
\lambda){\rm d}v, \;\;\;\;\nu=-\frac{4\pi q}{mc^{2}},
$$
$$
\varphi\mid_{\partial D}=-\frac{2\alpha q}{m}u_{01}, \;\;\;\psi\mid_{\partial D}=
\frac{q}{mc}u_{02} \eqno(6)
$$
in the subspace
$$
(\partial_{r}\varphi_{i},d_{i})=0, \;\;\;(\partial_{r}\psi_{i},d_{i})=0,\;\;
i=1,\ldots, N.   \eqno(7)
$$
Then the VM system has the solution
$$
E=\frac{m}{2\alpha q}\partial_{r}\varphi, \;\;\;B=\frac{d}{d^{2}}
(\beta+\int^{1}_{0}(d\times J(tr), r)\;{\rm d}t)-[d\times\partial_{r}\psi]
\frac{mc}{qd^{2}}, \eqno(8)
$$
where
$$
J\stackrel{\triangle}{=}\frac{4\pi}{c}\sum_{k=1}^{N}q_{k}\int_{R^{3}}vf_{k}\;{\rm d}v, \;\;\beta- const.
$$
To this solution there correspond potentials
$$
U=-\frac{m}{2\alpha q}\varphi, \;\;\;A=\frac{mc}{qd^{2}}\psi d+A_{1}(r),\;\;\;
(A_{1},d)=0 \eqno(9)
$$
with conditions (2).}

The proof see in [6].

We introduce the notations
$$
j_{i}=\int_{R^{3}}vf_{i}\;{\rm d}v, \;\;\;\; \rho_{i}=\int_{R^{3}}f_{i}\;{\rm d}v,\;\;\;i=1,\ldots,N.
$$
and impose the following condition:

{\bf B.} There exist  vectors $\beta_{i}\in R^{3}$ such that $j_{i}=\beta_{i}\rho_{i}$, $i=1,\ldots,N$.

Let condition {\bf B} holds, then system (5) becomes
$$
\bigtriangleup\varphi=a(\lambda)\mu\sum_{i=1}^{N}q_{i}A_{i},\;\;\;\;\;\;\;\;
$$
$$
\eqno(10)
$$
$$
\bigtriangleup\psi=a(\lambda)\nu\sum_{i=1}^{N}q_{i}(\beta_{i}, d)A_{i},
$$
where
$$
A_{i}(\lambda,l_{i}\varphi, k_{i}\psi)\stackrel{\triangle}{=}\int_{R^{3}}\hat{f_{i}}\;
{\rm d}v. \eqno(11)
$$

From now on, for simplicity, we consider the auxiliary vector $d$ directed
along the axis $Z$. Then $\varphi=\varphi(x,y)$, $\psi=\psi(x,y)$,
$x,y\in D\subset R^{2}$ in system (10). Moreover, let $N\geq 3$ and
$\frac{k_{i}}{l_{i}}\neq const$.

Let $D$ be a bounded domain in $R^{2}$ with boundary $\partial D$ of class $C^{2,\alpha}$,
$\alpha\in(0,1]$.

We introduce a continuous vector-function of parameter $\lambda$
$$
\varepsilon(\lambda)=(l_{1}\varphi^{0}, k_{1}\psi^{0}, \alpha_{1}, d_{1},
\ldots, l_{N}\varphi^{0}, k_{N}\psi^{0},\alpha_{N}, d_{N})\in R^{4N} \eqno(12)
$$
and a contraction of the set (the value of this vector) induced by the vector
(12) and the boundary conditions (3) for the local densities of the charge and the
current
\vspace{0.4cm}
$$
\Omega=\left\{\varepsilon\mid
\begin{array}{l}
\sum_{k=1}^{N}q_{k}A_{k}(l_{k}\varphi^{0}, k_{k}\psi^{0},\alpha_{k},d_{k})=0\\[0.4cm]
\sum_{k=1}^{N}q_{k}(\beta_{k},d)A_{k}(l_{k}\varphi^{0},k_{k}\psi^{0},
\alpha_{k},d_{k})=0
\end{array}
\right\} \eqno(13)
$$
with
$$
\varphi^{0}=-\frac{2\alpha q}{m}u_{01}, \;\;\;\psi^{0}=\frac{q}{mc}u_{02},\;\; N\geq 3.
\eqno(14)
$$
We introduce the condition.

{\bf C}. Let $\varepsilon(\lambda)\in \Omega$ for $\forall\lambda$ from the some
open set of real axis for $x\in (-r,r)$.

Then the system (10) with the boundary conditions
$$
\varphi\mid_{\partial D}=\varphi^{0},\;\;\;\psi\mid_{\partial D}=\psi^{0} \eqno(15)
$$
has the trivial solution $\varphi=\varphi^{0}$, $\psi=\psi^{0}$ for
$\forall\lambda\in (-r,r)$ by (13), (14).By Theorem 1, the boundary-value
problem (1), (2), (3) has the trivial solution
$$
E^{0}=\frac{m}{2\alpha q}\partial_{r}\varphi^{0}=0, \;\;\;r\in D\subset R^{2}, \;\;
B^{0}=\beta d_{1},
$$
$$
f^{0}_{i}=a(\lambda)\hat{f_{i}}(-\alpha_{i}(\lambda)v^{2}+c_{1i}+l_{i}\varphi^{0}
(\lambda),\; (v,d_{i}(\lambda)+c_{2i}+k_{i}\psi^{0}(\lambda)).
$$
for all $\lambda\in (-r,r)$. Moreover, $\rho^{0}\equiv 0$ and $j^{0}=0$
in a domain $D$.

{\bf Definition 1.} A point $\varepsilon_{0}=\varepsilon(\lambda_{0})\in\Omega$
is called {\it a bifurcation point} of solution of the VM system (I), (2), (3)
with {\it a bifurcation direction} $\varepsilon=\varepsilon(\lambda)$, where
$\lambda:(-r,r)\rightarrow\Omega$ be the continuous vector-function of $\lambda$,
if every neighborhood of the vector $(\varepsilon^{0},E^{0},B^{0},f^{0})$
corresponding to the trivial solution with $\rho^{0}=0$, $j^{0}=0$ in a
domain $D$ contains a vector $(\varepsilon^{*},E,B,f)$ with $\varepsilon^{*}=
\varepsilon^{*}(\lambda)$, $\lambda\in(-r,r)$ satisfying system (I) with
conditions (2), (3) such that
$$
\parallel E-E^{0}\parallel+\parallel B-B^{0}\parallel+\parallel f-f^{0}\parallel>0.
$$
Moreover, the densities $\rho$ and $j$ interior to domain $D$ need not vanish.

Using the Taylor expansion and singling out the linear terms, we rewrite  system
(10) in operator form as follows:
$$
[L_{0}-a(\lambda) L_{1}(\varepsilon(\lambda))]u-R(\varepsilon,u)=0, \eqno(16)
$$
where $\varepsilon=\varepsilon(\lambda)$ be a {\it bifurcation direction} of (12).
Here
$$
L_{0}=
\left[ \begin{array}{cc}
\bigtriangleup & 0 \\ [0.1cm]
0 & \bigtriangleup
\end{array}
\right],\;\;\;\; u=(\varphi-\varphi^{0}),\; \psi-\psi^{0})^{/}
$$
\vspace{0.7cm}
$$
L_{1}=\sum_{s=1}^{N}q_{s}
\left[ \begin{array} {ll}
\mu l_{s}\frac {\partial A_{s}} {\partial x} & \mu k_{s}\frac{\partial A_{s}}{\partial y}
\\ [0.6cm]
\nu l_{s}(\beta_{s},d)\frac {\partial A_{s}} {\partial x} &
\nu k_{s}(\beta_{s},d)\frac {\partial A_{s}} {\partial y}
\end{array}
\right]_{x=l_{s}\varphi^{0},\;y=k_{s}\psi^{0}}\stackrel{\triangle}{=}
\left[ \begin{array} {cc}
\mu T_{1} & \mu T_{2} \\ [0.5cm]
\nu T_{3} & \nu T_{4}
\end{array}
\right]. \eqno(17)
$$

Operator $R(\varepsilon(\lambda),u)$ is analytic in neighborhood of the point
$u=0$:
\vspace{0.5cm}
$$
R(\varepsilon,u)=\sum_{i\geq l}^{\infty}\sum_{s=1}^{n}\varrho_{is}(u)b_{s},
$$
where
$$
\varrho_{is}(u)\stackrel{\triangle}{=}\frac{q_{s}}{i!}(l_{s}u_{1}\frac{\partial}{\partial x}+k_{s}u_{2}
\frac{\partial}{\partial y})^{i}A_{s}(l_{s}\varphi^{0},\;k_{s}\psi^{0})
$$
are $i$th-order homogeneous forms in $u$ and
$$
\frac{\partial^{i_{1}+i_{2}}}{\partial x^{i_{1}}\partial y^{i_{2}}}A_{s}(\varepsilon,x,y)\mid_{x=l_{s}\varphi^{0},\;
y=k_{s}\psi^{0}}=0\;\;{\rm for}\;\;2\leq i_{1}+i_{2}\leq l-1,\;\;s=1,\ldots,N,
$$
$$
b_{s}\stackrel{\triangle}{=}(\mu, \;\nu(\beta_{s},d))^{/}.
$$

The existence problems for a bifurcation point, $\varepsilon\in\Omega$, of
the boundary-value problem (10), (15) can be restated as the existence problem
for a bifurcation point for the operator equation (16).

Let us introduce the Banach spaces $C^{2,\alpha}(\bar{D})$ and $C^{0,\alpha}(\bar{D})$
with the norms $\parallel\cdot\parallel_{2,\alpha}$ and
$\parallel\cdot\parallel_{0,\alpha}$, respectively, and let $W^{2,2}(D)$ be the
ordinary Sobolev  $L^{2}$- space in $D$.

Let us introduce the Banach space $E$ of vectors $u\stackrel{\triangle}{=}(u_{1},u_{2})^{/}$,
where $u_{i}\in L_{2}(D)$; $L_{2}$ is the real Hilbert space with inner
product $(\;,\;)$ and the corresponding norm$\parallel\cdot\parallel_
{L_{2}(D)}$. We define the domain $D(L_{0})$ as the set of vectors
$u\stackrel{\triangle}{=}(u_{1},u_{2})^{/}$with
$u_{i}\in \stackrel{\circ}{W^{2,2}}(D)$. Here $\stackrel{\circ}{W^{2,2}}(D)$ consists
of $W^{2,2}$ functions with zero trace on $\partial D$.

Then $L_{0}:\;D\subset E\longrightarrow E$ is a linear self-adjoint operator.
By virtue of the embedding
$$
W^{2,2}(D)\subset C^{0,\alpha}(\bar{D}),\;\;0<\alpha<1, \eqno(18)
$$
the operator $R: W^{2,2}\subset E\longrightarrow E$ is analytic in a neighborhood of the origin.
By the embedding (18), any solution of Eq. (16) in $D(L_{0})$ is a H$\ddot{o}$lder
function. Moreover, since the coefficients of
system (16) are constant, the vector $R(\varepsilon,u)$ is analytic, and $\partial D\in C^{2,\alpha}$;
it follows from well-known results of regularity theory for weak solutions of
elliptic equations [8] that the generalized solutions of Eq. (16) in
$\stackrel{\circ}{W^{2,2}}(D)$ actually belong to $C^{2,\alpha}(\bar{D})$.

The operator $L_{1}\in L(E\rightarrow E)$ is bounded and linear. Under the above
assumptions on $L_{0}$ and $L_{1}$, all singular points of the operator
$$
L(\varepsilon)\stackrel{\triangle}{=}L_{0}-a(\lambda)L_{1}(\varepsilon(\lambda))
$$
are Fredholm points in sense [9].

Let us introduce the conditions

{\bf I.} $T_{1}<0$;

{\bf II.} $T_{1}T_{4}-T_{2}T_{3}>0$.

If $\frac{\partial\hat{f}_{k}}{\partial x}\mid_{x=y=0}>0$, then inequality
{\bf II} is satisfied. Let us introduce the matrix $\;\;$ $\parallel\Theta_{ij}
\parallel_{i,j=1,\ldots,N}$, where $\Theta_{ij}=q_{i}q_{j}(l_{j}k_{i}-
k_{j}l_{i})(\beta_{j}-\beta_{i},d)$.
If the derivatives $\frac{\partial A_{i}}{\partial x}$, $\frac{\partial A_{i}}
{\partial y}$ are positive and equal at the point $x=y=0$,
$\Theta_{ij}>0$, $i\neq j$, then conditions {\bf I}, {\bf II} are satisfied.
Evidently, the elements of $\Theta_{ij}$ are nonnegative, because of identities
$$
sign\frac{q_{i}}{l_{i}}=sign q, \;\;\frac{(d_{i},d)}{\alpha_{i}}=\frac{d^{2}}
{\alpha}\frac{k_{i}}{l_{i}},\;\;q<0, \;\;q_{i}>0, \;\;i=2,\ldots,N.
$$
Let us denote by $\Xi$ the matrix generating the operator $L_{1}$.

{\bf Lemma 1.} Let  conditions {\bf I}, {\bf II} be satisfied. Then the matrix
$\Xi$ in (17) has two simple eigenvalues
$$
\chi_{+}(\varepsilon)=\mu T_{1}+o(1), \mu=-\frac{8\pi\alpha\mid q\mid}{m},
$$
$$
\chi_{-}(\varepsilon)=\eta\frac{T_{1}T_{4}-T_{2}T_{3}}{T_{1}}\frac{1}{c^{2}}+
o(\epsilon), \;\;\;\eta=\frac{4\pi\mid q\mid}{m}>0
$$
for $\frac{1}{c^{2}}\rightarrow 0$. \\
The eigenvectors corresponding to negative eigenvalue $\chi_{-}$ of the matrices
$\Xi$ and $\Xi^{/}$, respectively, are
$$
\left[ \begin{array}{c}
c_{1} \\ [0.4cm]
c_{2}
\end{array}
\right]=
\left[ \begin{array} {c}
-\frac{T_{2}}{T_{1}} \\ [0.4cm]
1
\end{array}
\right]+O(\frac{1}{c^{2}}), \;\;\;\;
\left[ \begin{array}{c}
c^{*}_{1} \\ [0.4cm]
c^{*}_{2}
\end{array}
\right]=
\left[ \begin{array} {c}
0 \\ [0.4cm]
1
\end{array}
\right]+O(\frac{1}{c^{2}}). \eqno(19)
$$
Proof see in [6, Lemma 4].

Let $\mu_{0}$ is a n-multiple eigenvalue of Dirichlet problem
$$
-\bigtriangleup e=\mu e, \;\;\;\;e\mid_{\partial D}=0, \eqno(20)
$$

Let us introduce the condition:

{\bf D.} Suppose that condition {\bf C} is satisfied and there exists
$\lambda_{0}\in (-r,r)$ such that $a(\lambda_{0}\chi_{-}(\varepsilon(\lambda_{0}))+
\mu_{0}=0$; moreover, $a(\lambda)\chi_{-}(\varepsilon(\lambda))$ is the
monotone increasing (decreasing) function in neighborhood of $\lambda_{0}$.

Rewrite the system (16) in the form
$$
Bu-B_{1}(\lambda)u=R(\varepsilon(\lambda),u), \eqno(21)
$$
where we introduce the following notations
$$
B=L_{0}+\mu_{0},\;\;\;\;B_{1}(\lambda)=L_{1}(\varepsilon(\lambda))+\mu_{0}.
$$
The operator $B$ is Fredholm and it has $n$ eigenfunctions of the form
$$
{\bf e}_{i}=\left(
\begin{array}{c}
c_{1}\\[0.2cm]
c_{2}
\end{array}
\right)e_{i}, \;\;\;i=1,\ldots,n.
$$
Vector $(c_{1},c_{2})^{/}={\bf c}$ is defined by (19) up to the constant multiplier
and $L_{1}${\bf c}=$\chi_{-}${\bf c}. Therefore,
$$
B_{1}{\bf e}_{i}=a(\lambda)L_{1}(\varepsilon(\lambda_{0})){\bf c}e_{i}-
a(\lambda_{0})L_{1}(\varepsilon_{0}){\bf c}e_{i}=(a(\lambda)\chi_{-}
(\varepsilon(\lambda))-a(\lambda_{0}\chi_{-}(\varepsilon_{0})){\bf c}e_{i}
$$
and by taking into account condition {\bf D} we obtain the following identities
$$
<B_{1}{\bf e}_{i},{\bf e}_{j}>=(a(\lambda)\chi_{-}(\varepsilon(\lambda))+
\mu_{0})\mid {\bf c}\mid^{2}\delta_{ij},\;\;i,j=1,\ldots,n. \eqno(22)
$$

{\bf Theorem 2.} {\it Let conditions {\bf A}, {\bf B}, {\bf C}, {\bf D}, {\bf I},
{\bf II}, as well one of the following conditions be satisfied:\\
1)$\mu_{0}$ is odd-multiple eigenvalue of problem (20); \\
2) $f_{i}=a(\lambda)\hat{f}_{i}(-\alpha_{i}(v^{2}+\varphi_{i}+v\cdot d_{i}(\lambda)+
\psi_{i})$. Then $\varepsilon_{0}=\varepsilon(\lambda_{0})$ is a bifurcation point
of the boundary problem (1), (2), (3) with bifurcation direction $\varepsilon=
\varepsilon(\lambda)$}.

Proof. In order that to prove Theorem 2, we need to show that $\lambda_{0}$
is a bifurcation point of Eq.(21). The corresponding finite-dimensional
branching system was given in the paper [6, Eq.(5.1)] for the case, when $l_{i}$, $d_{i}$
are independent of $\lambda$ and $a(\lambda)=\lambda\in R^{+}$. It is shown in
the same place that $\lambda_{0}$ is a bifurcation point if and only if $\lambda_{0}$
is a bifurcation point of branching system.

In general case, when the values $\alpha_{i}(\lambda)$, $d_{i}=d_{i}(\lambda)$
are dependent of $\lambda$, we can reduce the bifurcating system  to the form
$$
(a(\lambda)\chi_{-}(\varepsilon(\lambda))+\mu_{0})(c_{1}^{2}+c_{2}^{2})\xi_{i}+
r_{i}(\xi_{1},\ldots,\xi_{n},\lambda)=0, \;\;i=1,\ldots,n. \eqno(23)
$$
with $\mid r(\xi,\lambda)\mid=o(\mid\xi\mid)$, using identities (22) and  results
from [6]. Here the vector-function $r(\xi,\lambda)$ in equation (23) is potential
of $\xi$, if the condition 2) of Theorem 2 be satisfied.

Because the continuous function $a(\lambda)\chi_{-}(\xi(\lambda))+\mu_{0}$
is equal to zero at the point $\lambda_{0}$, and be monotonic in it neighborhood,
then in any neighborhood of point $\lambda_{0}$, $\xi_{0}=0$ exists a couple
$\lambda^{*}$, $\xi^{*}$ with $\xi^{*}\neq 0$ satisfying the system (23) [
6, see the proof of Lemma 8 and terminology in 10]. Hence, $\lambda_{0}$ is a bifurcation point of
the system (23). Theorem 2 is proved.


\begin{thebibliography} {}
\bibitem{1}
A.A. Vlasov. {\it Many-particle theory and its application to plasma.} Gordon
and Breach, engl. Auflage, 1961.
\bibitem{2}
G.A. Rudykh, N.A. Sidorov, A.V. Sinitsyn. {\it Stationary solutions of a
system of Vlasov-Maxwell equations}. Soviet Phys. Dokl., 33: 673-674, 1986.
\bibitem{3}
P. Braasch. {\it Semilineare elliptische Differentialgleichungen und das
Vlasov-Maxwell-System.} Dissertation an der Fakult$\ddot{a}$t f$\ddot{u}$r
Mathematik der Ludwig-Maximilians-Universit$\ddot{a}$t M$\ddot{u}$nchen.
M$\ddot{u}$nchen: Utz, Wiss., 1997.
\bibitem{4}
N.A. Sidorov, A.V. Sinitsyn. {\it On nontrivial solutions and bifurcation
points of the Vlasov-Maxwell system.} Dokl. Akad. Nauk of Russia, V.349, N 1,
26-28, 1996. /in Russian/
\bibitem{5}
N.A. Sidorov, A.V. Sinitsyn. {\it On bifurcation of solutions of the Vlasov-
Maxwell system}. Sibirskii Matemat. Zhurnal, V.37, N 6, 1367-1379, 1996. /in
Russian/
\bibitem{6}
N.A. Sidorov, A.V. Sinitsyn. {\it Analysis of Bifurcation Points and Nontrivial
Branches of solutions to the Stationary Vlasov-Maxwell System.} Mathematical
Notes, V.62, N.2, 223-243, 1997. /translated from Matematicheskie Zametki, V.62,
N 2, 268-292, 1997/.
\bibitem{7}
Yu.A. Markov, G.A. Rudykh, N.A. Sidorov, A.V. Sinitsyn and D.A. Tolstonogov. {\it
Steady-state solutions of the Vlasov-Maxwell system and their stability.}
Acta Appl. Math., 28: 253-293, 1992.
\bibitem{8}
O.A. Ladyzhenskaya, N.N. Uralzeva.  {\it Linear and nonlinear equations of elliptic type.} Moscow, Nauka,
1964. /in Russian/
\bibitem{9}
M.M. Vayenberg, V.A. Trenogin. {\it The theory of branching of solutions of
nonlinear equations.} Moscow, Nauka, 1969. /in Russian/
\bibitem{10}
N.A. Sidorov. {\it The general regularization questions in the problems of bifurcation theory.}
Irkutsk Gos. University, Irkutsk. 1982. /in Russian/


\end{thebibliography}
\end{document}